\documentclass[fleqn,usenatbib]{mnras}
\usepackage{newtxtext,newtxmath}
\usepackage{amsmath}
\usepackage[T1]{fontenc}
\DeclareRobustCommand{\VAN}[3]{#2}
\let\VANthebibliography\thebibliography
\def\thebibliography{\DeclareRobustCommand{\VAN}[3]{##3}\VANthebibliography}
\usepackage{graphicx}	
\usepackage{amsmath}

\title[Double degenerate system NLTT 16249]{The total mass of the close, double degenerate (DA+DQ)
system NLTT~16249}
\author[Vennes \& Kawka]{
St\'ephane Vennes $^{1}$\thanks{E-mail: svennes@iinet.net.au} and
Adela Kawka, $^{2}$\thanks{E-mail: adelakawka@gmail.com}\\
\\
$^{1}$Mathematical Sciences Institute, The Australian National University, ACT 0200, Australia\\
$^{2}$ Alexander Heights, WA 6064, Australia
}
\date{Accepted XXX. Received YYY; in original form ZZZ}
\pubyear{2024}
\begin{document}
\label{firstpage}
\pagerange{\pageref{firstpage}--\pageref{lastpage}}
\maketitle

\begin{abstract}
We revisit the binary and stellar properties of the double-degenerate system NLTT~16249. An analysis of new echelle spectra, supported by a joint study of a DQZ velocity template NLTT~44303, confirms the orbital period and constrains the mass ratio revealing a
carbon-polluted DQ white dwarf that is up to $\approx6$~percent more massive than its hydrogen-rich DA companion. Our new model atmosphere analysis of the DA and DQ components, constrained by an accurate Gaia parallax measurement that places the binary at a distance of 57.8~pc, reveals lower mass and temperature than previously estimated for both components, but with higher carbon and nitrogen abundances in the DQ atmosphere. The two components are nearly coeval and could have been generated following a single common envelope event.
\end{abstract}

\begin{keywords}
binaries: close --- binaries: spectroscopic --- stars: individual: NLTT~16249 --- individual: NLTT~44303 --- white dwarfs
\end{keywords}

\section{Introduction}
\label{introduction}
The double degenerate system NLTT~16249 is the only known close binary comprised of a carbon-polluted DQ white dwarf and a hydrogen-rich DA white dwarf \citep{ven2012a,ven2012b}. Although many close, hydrogen-rich pairs with a mass ratio close to unity are known \citep[see][]{deb2015, kaw2017, reb2017, nap2020, kil2020, mun2024}, even fewer heterogeneous pairs like the 1.17-d DA+DQ binary NLTT~16249 are known \citep[e.g., the DA+DB PG~1115+166:][]{max2002, ber2002} which may be simply due to the fact that the DQ phenomenon is itself relatively rare \citep{lim2015}. The chemical composition of the DQ white dwarf in NLTT~16249 is unique as well. It is the only known DQ white dwarf showing traces of nitrogen \citep{ven2012a}. The original abundance analysis of \citet{ven2012a} and based on the detection of C$_2$ and CN molecules constrained the abundance ratio to ${\rm C:N=50:1}$. The presence of nitrogen in the atmosphere of a DQ white dwarf is problematic. As a catalyst in the CNO-cycle, nitrogen is not expected to survive beyond the AGB phase unless nuclear burning in the shell above the core abruptly ceases, as postulated by \citet{ven2012b}, leaving traces of nitrogen that can be dredged-up to the atmosphere along with carbon as part of the DQ phenomenon \citep{duf2011}. The absence of any other trace constituents, such as calcium, in the atmosphere of the DQ (with the exception of carbon) or DA components appears to exclude an external origin to the nitrogen such as accretion from a debris disc.

We present a detailed analysis of the binary orbit and stellar properties of the double degenerate NLTT~16249 based on VLT/X-shooter echelle spectra and Gaia astrometric measurements (Section~\ref{observ}). We ascertain the effect of pressure shift on molecular bands with a joint analysis of the molecular bands and {Ca}~{\sc ii}~H\&K spectral lines in the polluted DQZ white dwarf NLTT~44303 (Section~\ref{sec-DQZ}). Next, we update the orbital parameters (Section~\ref{sec-orbit}), and revise the stellar parameters and the chemical composition of the DQ white dwarf (Section~\ref{sec-param}). We proceed with a summary and discussion revisiting the possible origin of this binary and the nature of the peculiar DQ atmosphere (Section~\ref{sec-final}).

\section{Observations}\label{observ}

\begin{figure}
\includegraphics[viewport=0 0 600 575, clip, width=1.0\columnwidth]{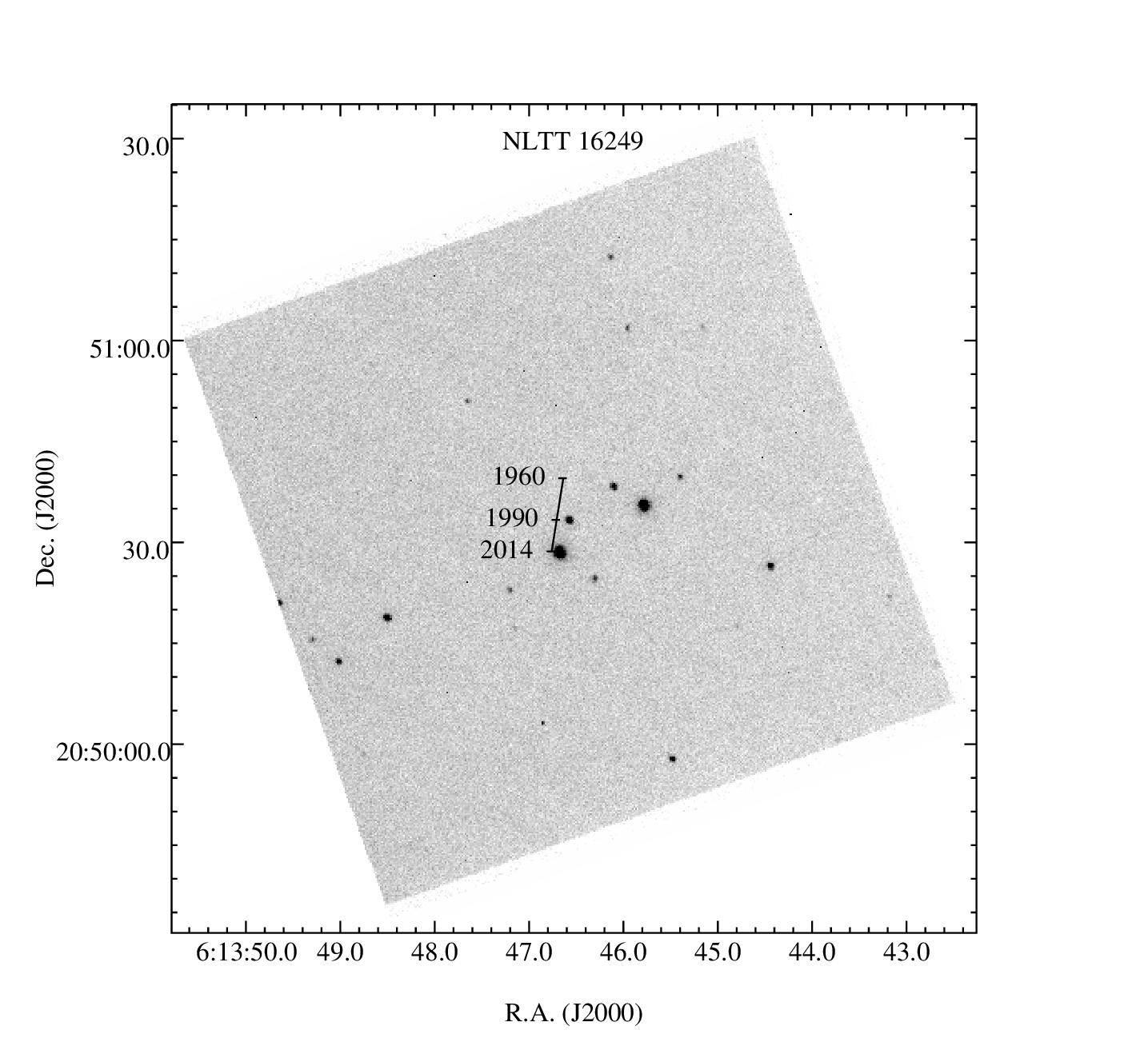}
\caption{VLT/X-shooter acquisition image (epoch $\approx$2014) showing the $\approx 1.5'\times1.5'$ field surrounding NLTT~16249.
The star is shown at the 2014 position and earlier positions (1960 and 1990) are estimated using $\mu_\alpha\cos{\delta}=+0.02$ and $\mu_\delta=-0.18''$~yr$^{-1}$.}
\label{fig-acq}
\end{figure}

We present new spectroscopic observations of the double degenerate system
NLTT~16249. We collected photometric data from published catalogues and eight ESO/VLT acquisition
images (see Fig.~\ref{fig-acq}). Most acquisition images were obtained using the V filter, while 
on UT 2014 January 21 the images were obtained using the R filter. Archival spectra used in the analysis of the DQZ white dwarf NLTT~44303 are also presented.

\begin{table}
\centering
\caption{X-shooter Spectroscopy}
\label{tbl-log}
\begin{tabular}{cccc}
\hline
UT date    &UT time (middle)&$t_{\rm exp}$& Arm \\
           &                &  (s)        &     \\
\hline
2010 11 07 &   07 53 56 &     2400 & UVB \\     
           &   07 54 02 &     2400 & VIS \\     
           &   07 54 05 & $4\times600$ & NIR \\
2013 10 31 &   07 44 29 &     3000 & UVB \\     
           &   07 44 04 &     2940 & VIS \\     
           &   07 44 37 & $5\times600$ & NIR \\
2013 11 30 &   04 59 55 &     3000 & UVB \\     
           &   04 59 30 &     2940 & VIS \\     
           &   05 00 03 & $5\times600$ & NIR \\
2014 01 07 &   03 20 06 &     3000 & UVB \\     
           &   03 19 41 &     2940 & VIS \\     
           &   03 20 14 & $5\times600$ & NIR \\
2014 01 21 &   05 45 39 &     3000 & UVB \\     
           &   05 45 14 &     2940 & VIS \\     
           &   05 45 47 & $5\times600$ & NIR \\
2014 01 30 &   02 14 29 &     3000 & UVB \\     
           &   02 14 05 &     2940 & VIS \\     
           &   02 14 38 & $5\times600$ & NIR \\
2014 01 30 &   03 30 26 &     3000 & UVB \\     
           &   03 30 01 &     2940 & VIS \\     
           &   03 30 34 & $5\times600$ & NIR \\
2014 01 31 &   02 12 02 &     3000 & UVB \\     
           &   02 11 37 &     2940 & VIS \\     
	   &   02 12 11 & $5\times600$ & NIR \\
\hline
\end{tabular}
\end{table}

\subsection{Spectroscopy}

We followed up on our initial observations of NLTT~16249 with seven sets of echelle spectra 
using the X-shooter spectrograph \citep{ver2011}
attached to the UT3 at Paranal Observatory between UT 2013 October 31 and 2014 
January 31 (Table~\ref{tbl-log}).
The slit-width was set to 0.5, 0.9 and 0.6 arcsec for the UVB, 
VIS and NIR arms, respectively. This setup provided a resolving power $R=\lambda/\Delta\lambda=$9900, 7450
and 7780 for the UVB, VIS and NIR arms, respectively, and a near-continuous wavelength coverage between 0.3 and 2.5~$\mu$m. The exposure times for
the UVB and VIS arms were 2940 and 3000 s, respectively, and for the NIR arm
we obtained five exposures of 600 s each. All observations were conducted at the parallactic angle.

Our attention was drawn to NLTT~44303 during observations at the Cerro Tololo Inter-American Observatory (CTIO). The spectra were obtained on UT 2008 February 24 using the RC-spectrograph and the low-dispersion grating KPGL2 and the order blocking filter WG360 delivering a resolution of 5-6\AA. The spectra showed the Ca{\sc ii}~H\&K doublet along with the C$_2$ Swan bands which should prove useful in estimating pressure shifts of the Swan bands observed in DQ white dwarfs.

We obtained a series of UV-Visual Echelle Spectrograph (UVES) spectra of the white dwarf NLTT~44303 from the ESO archives with the program IDs 165.H-0588(A), 69.D-0534(A), and 71.D-0383(A). The programs were designed to observe and detect new close double degenerate systems \citep[see][]{nap2020}. The series consist of 10 spectra each from the red and blue sides of the spectrograph in UT 2000 July 6 and 18, 2002 June 5, 2003 March 18 to 21, and 2003 June 22. Three of the spectra were poorly exposed and were excluded from the analysis. 

\subsection{Photometry and spectral energy distribution}

\begin{table}
\caption{Photometric and astrometric measurements}
\label{tbl-phot}
\begin{tabular}{ccccc}
\hline
Band     & $<\lambda>$ & magnitude &  Epoch     & Ref. \\
         &  $\mu$m   & (mag) &            &        \\
\hline
$U$      &  0.36     & 15.35 & $\approx$1960 & 1       \\
$B$      &  0.44     & 16.01 & $\approx$1960 & 1       \\
$V$      &  0.55     & 15.77 & $\approx$1960 & 1       \\
$r$      &  0.62     & 15.66$\pm$0.01& 2005.9 & 2      \\
$i$      &  0.76     & 15.51$\pm$0.01& 2005.9 & 2      \\
$g$      &  0.4866   & 15.886$\pm$0.006& 2011.6 & 3    \\
$r$      &  0.6215   & 15.874$\pm$0.004& 2011.6 & 3    \\
$i$      &  0.7545   & 15.949$\pm$0.006& 2011.6 & 3    \\
$z$      &  0.8679   & 16.076$\pm$0.004& 2011.6 & 3    \\
$y$      &  0.9633   & 16.173$\pm$0.003& 2011.6 & 3    \\
$J$      &  1.235    & 15.554$\pm$0.003& 2010.2   & 4    \\
$H$      &  1.662    & 15.496$\pm$0.005& 2010.2   & 4    \\
$K$      &  2.159    & 15.485$\pm$0.012& 2010.2   & 4    \\
$J\,^a$      &  1.235    & 14.87$\pm$0.04& 1997.9 & 5   \\
$H\,^a$      &  1.662    & 14.71$\pm$0.06& 1997.9 & 5   \\
$K\,^a$      &  2.159    & 14.61$\pm$0.08& 1997.9 & 5   \\
$W1\,^a$     &  3.353    & 15.39$\pm$0.02& 2010-2017 & 6   \\
$W2\,^a$     &  4.603    & 15.73$\pm$0.07& 2010-2017 & 6   \\
$W1\,^a$     &  3.353    & 15.80$\pm$0.04& 2010-2018 & 7   \\
$W2\,^a$     &  4.603    & 15.89$\pm$0.07& 2010-2018 & 7   \\
$G$     &  0.6405    & 15.799$\pm$0.003 & 2014.5-2017.4 & 8 \\
$G_{BP}$ & 0.5131    & 15.887$\pm$0.004 & 2014.5-2017.4 & 8 \\
$G_{RP}$ & 0.7778    & 15.625$\pm$0.005 & 2014.5-2017.4 & 8 \\
         &            &                 &           & \\
\multicolumn{2}{c}{Astrometry} & & Epoch & Ref.  \\
\hline
\multicolumn{2}{c}{$\pi$ (mas)} & 17.25$\pm$0.06 & 2014.5-2017.4 & 8 \\
\multicolumn{2}{c}{$\mu_\alpha\cos\delta$} (mas\,yr$^{-1}$) &  18.45$\pm$0.05  & 2014.5-2017.4 & 8  \\
\multicolumn{2}{c}{$\mu_\delta$} (mas\,yr$^{-1}$) & $-$179.5$\pm$0.04 & 2014.5-2017.4 & 8  \\
\hline
\end{tabular}\\
$^a$ Blended with a nearby star.\\
References: (1) \citet{egg1968}; (2) IPHAS2 \citep{bar2014}; (3) Pan-STARRS \citep{cha2016}; (4) UKIDSS GPS \citep{luc2008};
(5) 2MASS \citep{skr2006}; (6) unWISE \citep{sch2019}; (7) CatWISE2020 \citep{mar2021}; (8) Gaia EDR3 \citep{gai2021}.
\end{table}

Table~\ref{tbl-phot} lists photometric measurements of NLTT~16249 in the optical and infrared. The object
is in the Galactic plane and the GALEX sky surveys did not cover that location.
Most measurements are provided in the Vega system while the Pan-STARRS are provided in the AB system. A precise epoch for the $UBV$ measurements could not
be obtained although it was most certainly in the early part of the 1960 decade.
The measurements from the UKIDSS Galactic Plane Survey (GPS) were obtained from the WFCAM Science Archive \footnote{wsa.roe.ac.uk}: we list the weighted average of two separate entries. Table~\ref{tbl-phot} includes Gaia astrometric and photometric measurements which constrain the distance to the binary system to $57.8\pm0.2$ pc following \citet{bai2021}. The parallax given in \citet{leg2018}, $\pi=17.77\pm0.31$ is marginally consistent with the Gaia parallax.

We collected photometric observations obtained by the {\it Transiting Exoplanet Survey Satellite} \citep[TESS,][]{ric2015} in sectors
43 (2021 September 16 to October 12), 44 (2021 October 12 to November 6), 45 (2021 November 6 to December 2), 71 (2023 October 16 to November 11) and 72 (2023 November 11 to December 7). The wavelength coverage of the {\it TESS} bandpass is from 6000 to 10\,000 \AA. 
We used the 20 second cadence observations of NLTT~16249 (TIC 46048105) to search for possible eclipses.

\begin{figure}
\includegraphics[viewport=0 30 550 580, clip, width=1.0\columnwidth]{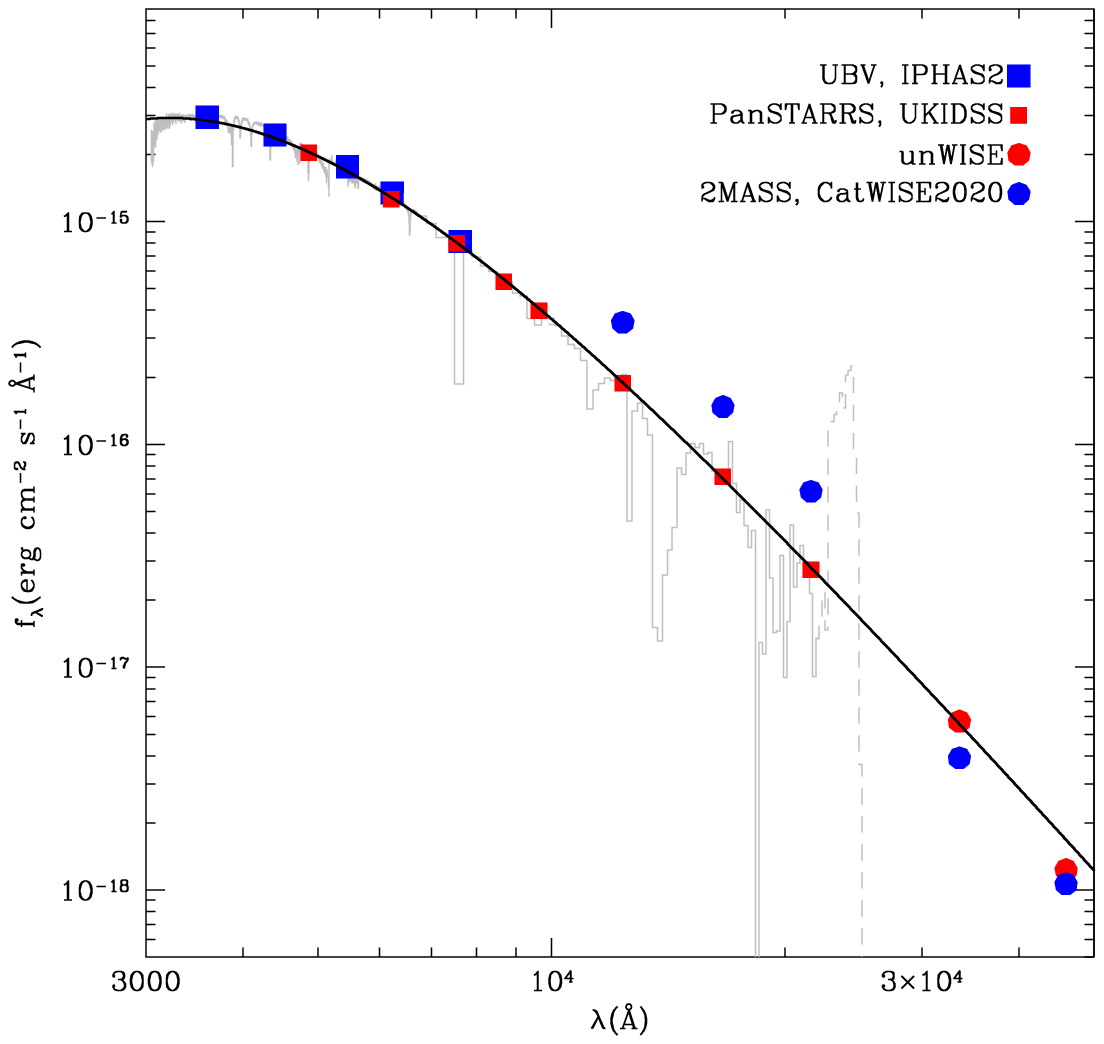}
\caption{Spectral energy distribution of the binary NLTT~16249 built using
X-shooter UVB, VIS, and NIR segments (grey lines) and photometric measurements (see Table~\ref{tbl-phot}) compared to a blackbody flux distribution at 9\,000~K (black line). The spectra are binned (100\AA) longward of
6\,800\AA\ and part of the NIR spectrum was rejected as unreliable (dashed line). The 2MASS (1-2 $\mu$m) and WISE (unWISE, CatWISE2020, 3-5 $\mu$m) show evidence of contamination from crowded stars (see text).}
\label{fig-sed}
\end{figure}

Fig.~\ref{fig-acq} shows the crowded field surrounding NLTT~16249. The spectral energy distribution (Fig.~\ref{fig-sed}) shows evidence of contamination in some IR measurements.
X-shooter NIR, VIS, and UVB spectra were co-added and their average scaled by factors of 2.01, 1.61, and 2.82, respectively. 
These figures correspond approximately to slit transmission factors ($T$) estimated
using the error function $T = {\rm erf}(w/\sigma)$, where $w$ is the slit width
and $\sigma=FWHM/0.833$ is the Gaussian parameter expressed in terms of the measured
full-width at half-maximum ($FWHM$) of the seeing disc, with both widths expressed in
units of arcsecond.
The IPHAS2 catalog lists a red object located 3.3$''$~N and 1.4$''$~W
relative to NLTT~16249 circa 2006. This nearby object is apparent in the VLT/X-shooter acquisition image shown in Fig~\ref{fig-acq}. The same object would have overlapped with NLTT~16269 in the 2MASS images obtained circa 1994. 
The WISE images obtained years later, circa 2010-2018, would still blend the two objects due to the broad WISE point-spread-function ($6^{\prime\prime}$).
The two sets of measurements, unWISE and CatWISE2020, benefited from improved processing of crowded areas such as that of NLTT~16249. However, variations between the two catalogs in the resulting flux measurements and source locations, as well as slight deviations from a blackbody flux distribution, indicate that systematic errors in the deblending procedure dominate the error budget. The UKIDSS infrared images show NLTT~16249 well separated from all nearby stars and these IR photometric measurements are uncontaminated.

\section{Analysis}

First, we analyse the calcium and molecular carbon spectra of the DQZ NLTT~44303 and establish a reliable velocity scale for Swan bands under pressure (Section~\ref{sec-DQZ}). Next we proceed with a radial velocity analysis of the DA and DQ components of the binary NLTT~16249 (Section~\ref{sec-orbit}).
Finally, we revise the stellar parameters of both binary components and, in particular, the photospheric composition of the DQ component (Section~\ref{sec-param}).

\subsection{A DQZ velocity template: molecular carbon pressure shift}\label{sec-DQZ}

The simultaneous presence of carbon and calcium in the atmosphere of NLTT~44303 allows us to
calibrate the effect of pressure on the wavelength position of the Swan bands \citep{ham1990,kow2010,blo2019}.

We fitted the Ca{\sc ii}~H\&K lines with Lorentzian functions.
The mean and dispersion of all valid radial velocity measurements are $\varv=39.1$\ km\,s$^{-1}$ and $\sigma=2.7$\ km\,s$^{-1}$ when measuring the narrow CaH\&K line cores, while they are $\varv=33.4$\ km\,s$^{-1}$ and $\sigma=7.2$\ km\,s$^{-1}$ when measuring the shallow CaH\&K wings. We adopted the weighted mean of the two measurements:
\begin{math}
    \varv = 38.4\pm2.5\ {\rm km\, s^{-1}}.
\end{math}
The DQZ NLTT~44303 does not show evidence of orbital motion that would reveal the presence of a close companion. 
We co-added the spectra in the heliocentric velocity frame and the resulting spectrum reached a signal-to-noise ratio of  $\approx$260 near the Ca{\sc ii}~H\&K doublet and $\approx$340 near the C$_2$$\lambda$5165 band and a resolving power of $\approx$40\,000.

\begin{figure}
\includegraphics[viewport=20 30 550 550, clip, width=1.0\columnwidth]{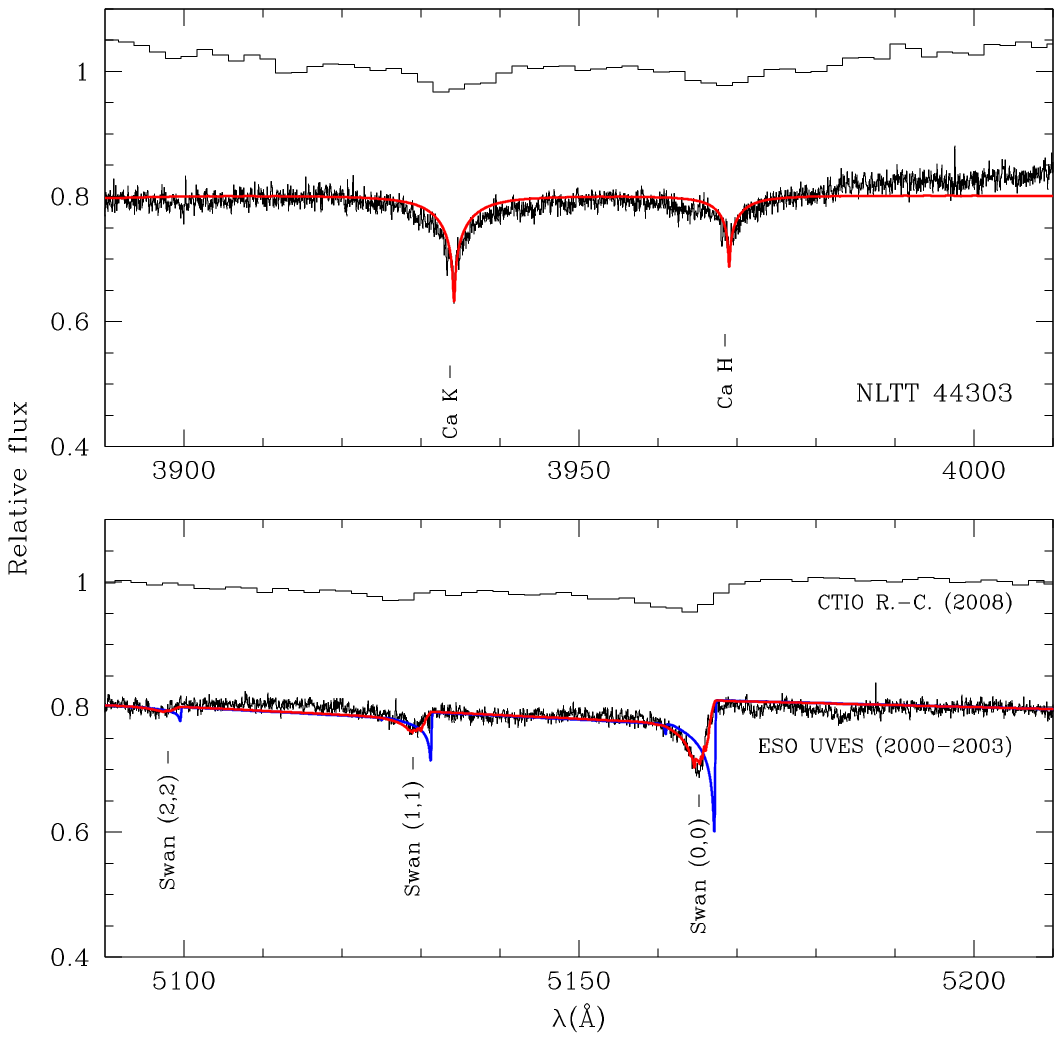}
\caption{CTIO and ESO spectra of the DQZ white dwarf NLTT~44303 showing the Ca{\sc ii}~H\&K doublet and C$_2$ Swan bands. The ESO/UVES spectra are fitted with model spectra (see text) including pressure shift (in red) and excluding it (in blue).}
\label{fig_44303}
\end{figure}

Fig.~\ref{fig_44303} shows the CTIO and ESO/UVES spectra covering the Ca{\sc ii} lines in the top panel and a set of C$_2$ Swan bands in the lower panel. We fitted the spectral energy distribution constrained by the Gaia parallax ($44.085\pm0.025''$) and obtained $T_{\rm eff}=9000\pm300$~K and $\log{g}=7.85\pm0.06$, making NLTT~44303 a slightly cooler and less massive white dwarf than estimated by \citet{sub2017} who measured $T_{\rm eff}=9280\pm320$~K and $\log{g/({\rm cm\,s^{-2}})}=7.93\pm0.06$. With the carbon abundance adjusted to $\log({\rm C/He})=-4.6\pm0.2$ and the calcium abundance estimated at $\log{\rm Ca/He}=-11.2\pm0.2$, we computed a detailed spectral synthesis of the C$_2$ Swan bands and Ca{\sc ii}~H\&K spectral lines.
The Swan bands were modelled in the ``just-overlapping line approximation'' \citep[see ][]{zei1982} with
improved band parameters for C$_2$ \citep{bro2013}. The Ca{\sc ii} lines were modelled with Lorentzian profiles including pressure shift and broadening parameters \citep{ham1975,bow1978,mon1986}.

We set the radial velocity of the spectral synthesis at the observed Ca{\sc ii}~H\&K velocity and proceeded with a calibration of the C$_2$ pressure shift in the energy levels.
The pressure shift in the C$_2$ lower electronic energy level $E_{\rm e}$ (in eV) at the ambient helium density $\rho_{\rm He}$ (in g\,cm$^{-3}$) is given by:
\begin{equation}
    \Delta E_{\rm e} = \alpha\, \rho_{\rm He},
\end{equation}
where $\alpha$ is a proportionality constant originally estimated at $\alpha=1.6$ by \citet{kow2010} for helium densities $\lesssim 0.25$~g\,cm$^{-3}$. However, a model atmosphere analysis of the DQp LHS~290 \citep{kow2010} requires a lower proportionality constant $\alpha=0.2$ unless the photosphere of LHS~290 is covertly made much thinner by the presence of hidden hydrogen. \citet{blo2019} rejected the presence of hydrogen and empirically evaluated the constant at $\alpha=0.2$ although no single value of $\alpha$ could reproduce the Swan bands of all peculiar DQ white dwarfs. The same expression given in wave number units ($T_{\rm e}$ in cm$^{-1}$) is:
\begin{equation}
    \Delta T_{\rm e} = C\, \rho_{\rm He},
\end{equation}
where $C = 8065.73\,\alpha$. We adjusted the parameter $\alpha$ using the UVES spectra of NLTT~44303 and obtained $\alpha=0.33$. Fig.~\ref{fig_44303} shows the best fit spectral synthesis at $\alpha=0.33$ compared to a spectral synthesis at $\alpha=0$. 
The pressure shift reconciles the C$_2$ and Ca{\sc ii}~H\&K radial velocities.

\subsection{Orbital parameters of the DA+DQ NLTT~16249}\label{sec-orbit}

\begin{table}
\centering
\caption{Radial velocity measurements (NLTT~16249)}
\label{tbl-vel}
\begin{tabular}{cccc}
\hline
HJD    & $\varv_{\rm DA}$ & HJD & $\varv_{\rm DQ}$  \\
 (2450000+)   &  (km\,s$^{-1}$) & (2450000+)   & (km\,s$^{-1}$)     \\
\hline
             &             &          \\
5507.83298  & 152.0 & 5507.83291 & $-$8.9\\
6596.82551  & 183.9 & 6596.82580 & $-$27.2\\
6626.71312  & $-$23.7 & 6626.71341 & 164.4 \\
6664.64418  &  64.4 & 6664.64447 & 74.7 \\
6678.74476  &  79.5 & 6678.74505 & 75.2 \\
6687.59766  &  31.0 & 6687.59794 & 109.2 \\ 
6687.65039  &  7.6  & 6687.65068 & 134.7 \\
6688.59589  & 122.9 & 6688.59617 & 23.3 \\
\hline
\end{tabular}
\end{table}

We revise the orbital parameters using all available ESO/X-shooter spectra and our new parametrized Swan profiles for the DQ component of the binary NLTT~16249.
For the DA white dwarf we measured velocity centroids of the H$\beta$ line core ($\pm2$\AA) in the UVB 
and of the H$\alpha$ line core in the VIS using Lorentzian functions. 
The DA velocities measured with the H$\alpha$ and H$\beta$ do not differ significantly, with $<\varv_{\rm H\beta}>-
<\varv_{\rm H\alpha}>=-1.0\pm5.9$\,km\,s$^{-1}$. Table~\ref{tbl-vel} lists the new DA velocities. Using all DA radial velocity measurements \citep[including][]{ven2012b} we revise the binary ephemeris:
\begin{displaymath}
P = 1.173677\pm0.000024\ {\rm d},
\end{displaymath}
\begin{displaymath}
T_0 ({\rm HJD})= 2456096.5843^{+0.0026}_{-0.0029}.
\end{displaymath}
The new ephemeris is more accurate and formally in agreement with that of \citet{ven2012b}.

For the DQ white dwarf we cross-correlated the spectra with one
another within wavelength ranges including CN (3820-3910 \AA) and C$_2$ molecular bands (4650-4750 and 5080-5180 \AA). See \ref{appenA} for other bands.
The absolute scale as opposed to a relative, cross-correlated scale, for the DQ white dwarf is more difficult to
establish. The tabulated band head values employed earlier \citep{ven2012a} do not mark the actual
centroids of the first series members, but rather a mid-point in the band opacity curves.

To establish the absolute velocity scale, we first cross-correlated the C$_2\,\lambda$5165 band with a spectral synthesis (see Section 3.3) with the parameter $\alpha=0.33$ as previously measured with the DQZ velocity template. 
The resulting velocity amplitudes are $K_{\rm DA} = 107.7\pm3.0$ and $K_{\rm DQ} = 101.7\pm3.4$~km\,s$^{-1}$ corresponding to a mass ratio
\begin{equation}
\frac{M_{\rm DQ}}{M_{\rm DA}} \equiv \frac{K_{\rm DA}}{K_{\rm DQ}} = 1.06\pm0.05.
\end{equation}
The systemic velocity uncorrected for the gravitational redshifts of the DQ and DA components, $z_{\rm DQ}$ and $z_{\rm DA}$, should only differ slightly, i.e., between $z_{\rm DQ}-z_{\rm DA}=1.9$ and $5.0$~km\,s$^{-1}$ for a mass ratio between 1.01 and 1.11 (see Table~\ref{tbl-param}, section 3.3). With the pressure shift parameter $\alpha=0.33$ the systemic velocity offset amounts to $-3.9$~km\,s$^{-1}$ while increasing the shift parameter to $\alpha=0.356\pm0.005$ provides uncorrected systemic velocities:
$\gamma_{\rm DQ}=74.2\pm2.1$ and $\gamma_{\rm DA}=70.8\pm1.9$~km\,s$^{-1}$. The offset $\gamma_{\rm DQ}-\gamma_{\rm DA}$ corresponds to the expected offset due to the effect of gravitational redshift. Variations in $\alpha$ parameters between NLTT~44303 and NLTT~16249 reflect uncertainties in current models for DQ white dwarfs. 
Revised kinematics based on the new systemic velocity corrected for gravitational redshifts are discussed in Section 4.
Table~\ref{tbl-vel} lists the corresponding DQ radial velocity measurements.

\begin{figure}
\includegraphics[viewport=0 20 550 550, clip, width=1.0\columnwidth]{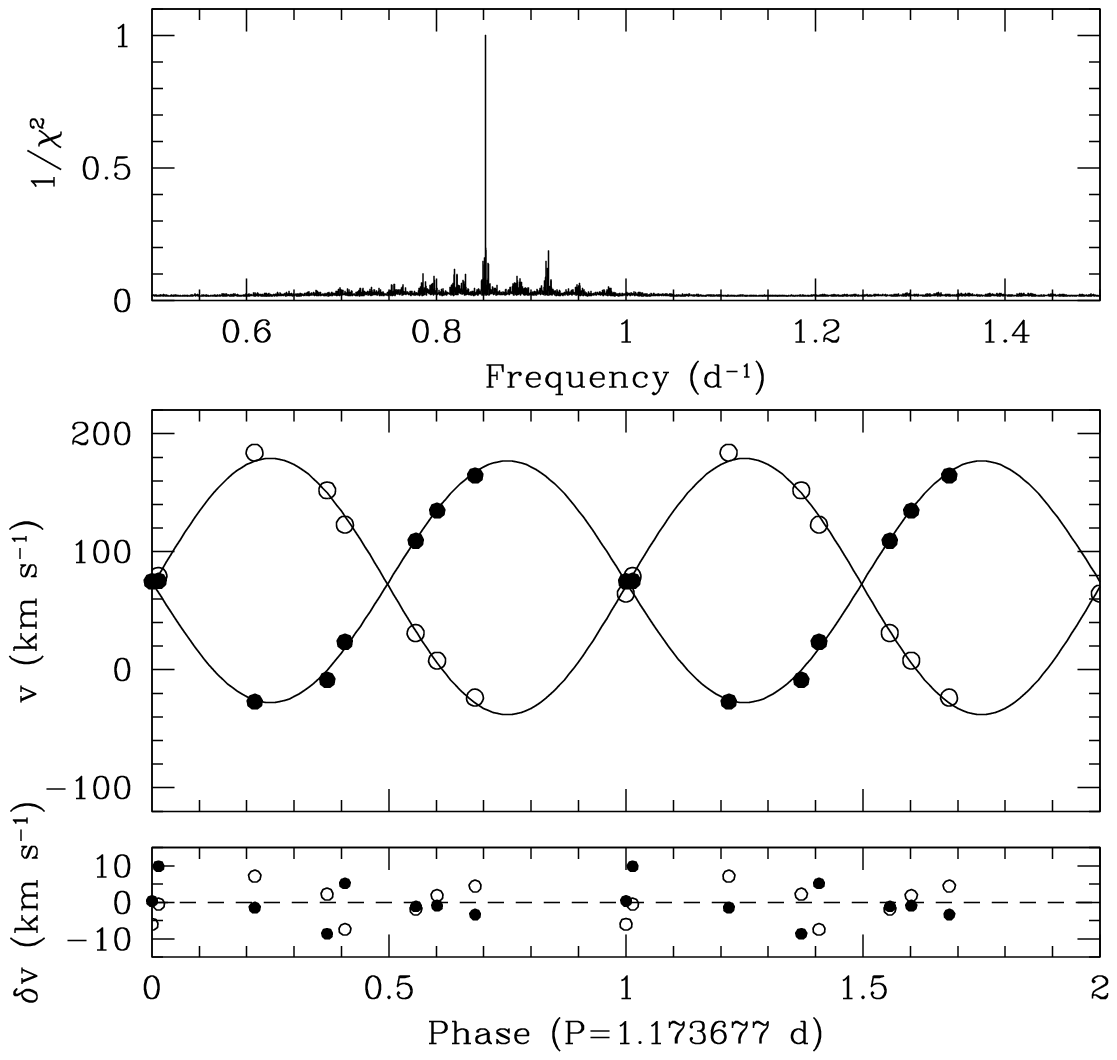}
\caption{(Top) Period analysis of all velocity measurements from \citet{ven2012b} and the new data revealing 
a unique solution confirming the period analysis of \citet{ven2012b}. (Middle) X-shooter radial velocity measurements phased on the adopted binary ephemeris.
The measurements for the DQ (full circles) and DA white dwarf (open circles) are compared to
circular orbits (full lines). (Bottom) the velocity residuals average 5.1~km\,s$^{-1}$ for the DQ 
white dwarf and 4.7~km\,s$^{-1}$ for the DA.}
\label{fig-phase}
\end{figure}

Fig.~\ref{fig-phase} shows the period analysis (top panel), phased radial velocity measurements for both components (middle panel), and the velocity residuals relative to circular orbits (bottom panel). The residuals of 5.1 and 4.7~km\,s$^{-1}$ for the DQ and DA components, respectively, are commensurate with expected radial velocity measurement errors at a resolving power $R\lesssim 10^4$.

The total mass of the system is constrained by the measured orbital period and velocities:
\begin{equation}
    M_{\rm DQ}+M_{\rm DA} = (K_{\rm DQ}+K_{\rm DA})^3\,\frac{P}{2\pi G\,\sin^3{i}},
\end{equation}
or, for any inclination angle:
\begin{equation}
    M_{\rm DQ}+M_{\rm DA} \ge (K_{\rm DQ}+K_{\rm DA})^3\,\frac{P}{2\pi G}=1.02\,{\rm M_\odot},
\end{equation}
and the projected separation is estimated at
\begin{equation}
    a\sin{i}=\frac{P}{2\pi}\,(K_{\rm DQ}+K_{\rm DA})=4.86\pm0.14\,R_\odot.
\end{equation}
The TESS data were obtained between BJD$=$2\,459\,474 and 2\,460\,285, some 9.25 years after the initial epoch of our spectroscopic data (see $T_0$ above). Accounting for the total relative velocity of the two bodies at conjunction an eclipse is expected to last at most 160~s or $\approx$8 bins of 20~s with a maximum depth of $\approx$50 percent. With a projected phase error of $\Delta \phi =\pm 0.066$ at the epoch of TESS data acquisition, the orbital ephemeris is not sufficiently accurate to phase the TESS data with an accuracy of 20~s, i.e., with a phase error of $\Delta \phi = 2\times10^{-4}$, but we would still expect eclipses, if any, to fall in the ranges $\phi=$0.43-0.57 and 0.93-0.07.
We scanned the data with orbital periods within a 2$\sigma$ range allowed by the orbital ephemeris using the spectroscopic initial epoch $T_0$. In order to build reliable phase bins of 20~s duration the orbital period must be scanned with a relative precision of $\Delta P/P \approx 2\times10^{-7}$. 

Combining all available TESS data we generated a series of 481 phased light curves with 5000 bins of $\approx20$~s covering the likely period range ($\pm2\sigma$) and we searched for maximum depth in each phased light curve. The light curves reached a signal-to-noise ratio of $\approx 50$, i.e., with features of average depth of 2 percent. We combined both eclipses, i.e., inferior and superior conjunctions, and increased the signal-to-noise ratio to 70, i.e., 1.4 percent average depth. We fitted each light curve with eclipse models of maximum depths ranging from 8 percent ($\approx$5 bins wide and lasting 100~s to zero percent. There were no eclipses deeper than 5 percent at 99 percent confidence level or 4 percent at 90 percent confidence level.
Inclinations above $89.73^\circ$ result in eclipses deeper than 5 percent and lasting at least 80~s (or $\approx$4 bins). The probability of an inclination larger than $89.73^\circ$ is only $P=0.5$ percent.

\subsection{Revised stellar parameters}\label{sec-param}

The present analysis takes advantage of refined constraints on the masses and radii of both binary components. To measure the stellar parameters, we followed these steps: First, we fixed the mass ratio at three consecutive values within the 1$\sigma$ range at $q=$1.01, 1.06, and 1.11~$M_\odot$. Next, we built a spectro-photometric flux template based on X-shooter spectra and photometric measurements listed in Table~\ref{tbl-phot}. The template was built by fitting to the photometric measurements the X-shooter spectra convolved with the photometric bandpasses and corrected using low-order polynomials. Then, we fitted to the template the total model flux in a four dimensional grid $(T_{\rm eff, DA}, g_{\rm DA}, T_{\rm DQ}, {\rm C/He})$:
\begin{equation}
    f_{\rm \lambda} = \frac{4\pi}{D^2}\ \Big{(} R^2_{\rm DA} H_{\rm \lambda, DA}+R^2_{\rm DQ} H_{\rm \lambda, DQ}\Big{)},
\end{equation}
where $D=57.8$~pc, $H_\lambda$ is the Eddington flux, and the radii $R_{\rm DA}$ and $R_{\rm DQ}$ are extracted from evolutionary tracks obtained from the Montr\'eal White Dwarf Database\footnote{https://www.montrealwhitedwarfdatabase.org/evolution.html} \citep[MWDD;][]{bed2020} and from \citet{ben1999}. The tracks allow for the conversion $(T_{\rm eff},g) \rightarrow (M,R,\tau_{\rm cool})$. For a given $T_{\rm eff}$, the quantities $M$, $R$, and $g$ are interchangeable. For a given $g_{\rm DA}$, the DQ surface gravity $g_{\rm DQ}$ is calculated from the mass ratio $q$.
The models $H_{\rm \lambda, DA}$ and $H_{\rm \lambda, DQ}$ suitable to NLTT~16249 have previously been described \citep{ven2012a}. As before, the line profiles were modelled in the ``just-overlapping line approximation'' \citep[see ][]{zei1982}, but with improved band parameters for C$_2$ \citep{bro2013} and CN \citep{bro2014}. Following our radial velocity analysis we adopted a pressure shift parameter $\alpha=0.356$ for both C$_2$ and CN band calculations. Carbon line opacities in the ultraviolet are included.

\begin{figure}
\includegraphics[viewport=0 0 560 560, clip, width=0.95\columnwidth]{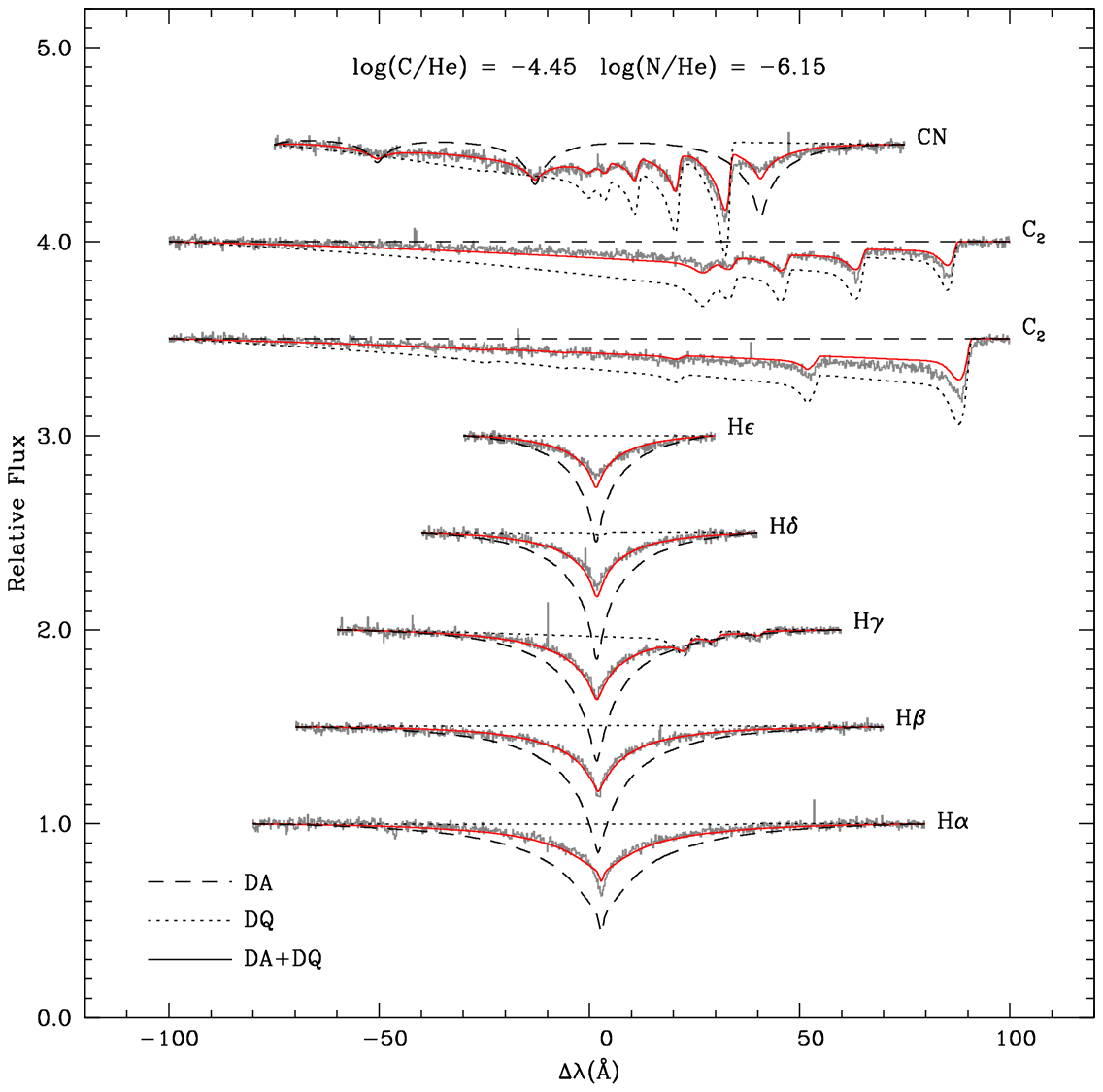}
\caption{X-shooter spectrum of NLTT~16249 (grey lines) and spectral decomposition using spectral syntheses of the DA white dwarf (short dashed lines), the DQ white dwarf (dotted lines), and the combined spectra (full lines).}
\label{fig_spec}
\end{figure}

Finally we computed two sets of $\chi^2$ values: The first based on the complete spectral range (absolute flux) from 3600 to 6800\AA\ and encompassing all line transitions of interest, and a second one restricted to normalized line profiles (relative flux). The sets of $\chi^2$ values were combined in seeking the best fit parameters. We adopted a very conservative 5 percent error on absolute flux measurements which reflect variations between different photometric systems. Table~\ref{tbl-param} lists the best fit parameters at three consecutive mass ratios in a range constrained by the orbital analysis with the best fit carbon and nitrogen abundances of $\log{\rm C/He}=-4.45\pm0.10$ and $\log{\rm N/He}=-6.15\pm0.10$. The total mass of the system is nearly constant at $M_{\rm DA}+M_{\rm DQ}=0.98\pm0.05\,M_\odot$ and marginally in agreement ($1\sigma$) with the minimum mass, $M_{\rm DA}+M_{\rm DQ}\gtrsim1.02\,M_\odot$, obtained from the orbital analysis. We conclude that the total mass of the system is very close to 1$M_\odot$ with nearly equal components, and that the inclination is close to 90$^\circ$. We note that the age overlap is closer at the prefered mass ratio ($q=1.01$). Fig.~\ref{fig_spec} shows the spectral decomposition for the best fit parameters (Table~\ref{tbl-param}) at $q=1.01$ and $\log{\rm C/He}=-4.45$. Minor flux residuals are apparent in the H$\alpha$ line core and in the first line (0,0) of the C$_2\,\lambda$5165 band. Results from earlier spectroscopic analyses that implied a much higher systemic mass and lower inclination \citep[e.g.,][]{ven2012a} should be discarded.

The absence of discernible eclipses in the TESS data imply $i< 89.73^\circ$, but we cannot exclude shallow, grazing eclipses ($89.67<i< 89.73^\circ$).
The mass function remains compatible (at $\approx1.5\sigma$ level) with the spectroscopic mass measurements down to an inclination of $\gtrsim80^\circ$.

Stellar parameter measurements may be affected by the presence of substantial reddening in the Galactic plane. However, at close distance ($\lesssim 100$~pc) the local cavity is relatively dust free and 2D emission maps ($E_{B-V}$) covering the full line-of-sight across the Milky Way \citep{sch1998} would overestimate the actual dust density toward nearby stars, e.g., $E_{B-V}=0.69$ in the case of NLTT~16249. The 3D extinction maps of \citet{lal2014} show a low dust density gradient toward NLTT~16249, i.e., at the bottom of the color scale in their Fig.~1, 
\begin{equation}
    \frac{d\,E_{B-V}}{d r}=1.0\times10^{-4}\ {\rm mag\, pc^{-1}}.
\end{equation}
At a distance of 57.8~pc the integrated extinction is constrained to $E_{B-V}\lesssim 0.006$~mag. Our series of X-shooter spectra impose an upper limit of 6~m\AA\ to the NaD1 equivalent width corresponding to a very low extinction coefficient $E_{B-V} << 0.01$~mag \citep{poz2012}. Assuming $E_{B-V}= 0.006$~mag, the fitted temperatures increase modestly, $\Delta T_{\rm eff,DA}\approx +30$~K and $\Delta T_{\rm eff,DQ}\approx +50$~K, while the individual masses remain unchanged.

\begin{table}
\centering
\caption{Stellar parameters (NLTT~16249) $\log{\rm C/He}=-4.45$}
\label{tbl-param}
\begin{tabular}{cccc}
\hline
       &      &  $q$ &       \\
\cline{2-4}
       & 1.01 & 1.06 & 1.11  \\
\hline
       DA      &        &  &  \\
$T_{\rm eff}$ (K) & 8132$\pm$50  & 8085$\pm$50 & 8044$\pm$50 \\
$\log{g}$ (cgs)   & 7.80$\pm$0.05  & 7.78$\pm$0.05 &  7.75$\pm$0.05\\
$M$ ($M_\odot$)   & 0.484$\pm$0.025  & 0.473$\pm$0.025 & 0.458$\pm$0.025 \\
$R$ ($10^{-2}\,R_\odot$)   & 1.45$\pm$0.04  & 1.47$\pm$0.04 & 1.49$\pm$0.04 \\
$z$ (km\,s$^{-1}$) &      21.2$\pm$1.7 &         20.5$\pm$1.7 & 19.6$\pm$1.6 \\
$\tau_{\rm cool}$ (Gyr)         & 0.83$\pm$0.04  & 0.83$\pm$0.04 & 0.81$\pm$0.04 \\
      DQ            &   &  &  \\
$T_{\rm eff}$ (K) & 8347$\pm$50  & 8368$\pm$50 & 8390$\pm$50 \\
$\log{g}$ (cgs)   & 7.86$\pm$0.05  & 7.88$\pm$0.05 & 7.90$\pm$0.05\\
$M$ ($M_\odot$)   & 0.497$\pm$0.025 & 0.508$\pm$0.025 & 0.519$\pm$0.025 \\
$R$ ($10^{-2}\,R_\odot$)   & 1.37$\pm$0.04  & 1.36$\pm$0.04 & 1.34$\pm$0.04 \\
$z$ (km\,s$^{-1}$) &      23.1$\pm$1.9 & 23.8$\pm$1.9 & 24.6$\pm$2.0 \\
$\tau_{\rm cool}$ (Gyr)         & 0.87$\pm$0.04  & 0.88$\pm$0.04 & 0.90$\pm$0.04 \\
\hline
\end{tabular}
\end{table}

\section{Summary and final thoughts}\label{sec-final}

The new constraints placed on the mass and radius of the two binary components by the Gaia parallax and the orbital mass ratio greatly improved our understanding of the double degenerate system NLTT~16249. The mass ratio is very close to unity and the individual masses close to $0.5\,M_\odot$ each. Moreover, their cooling ages are also very close which, along with nearly identical progenitor masses, imply that the objects are nearly coeval. However, one object survived with a hydrogen layer thick enough to have survived convective mixing ($q_{\rm H} = M_{\rm H}/M_* \gtrsim 10^{-5}$), while the other object lost its hydrogen envelope and displays a unique surface composition He:C:N$=1.4\times10^6$:50:1 not encountered in any other DQ white dwarfs. Our joint analysis of the DQZ white dwarf NLTT~44303 revealed a warm carbon-polluted atmosphere ($T_{\rm eff}=9\,000$~K) with evidence of external pollution ($\log{\rm Ca/He}$=-11.2). The presence of heavy elements in the atmosphere of DQ white dwarfs is relatively rare \citep{far2024}, a phenomenon attributed by \citet{blo2022} to the suppression of the Swan bands in the thin atmosphere of highly polluted white dwarfs.

The low mass of the DQ white dwarf in the binary NLTT~16249 confirms earlier analyses of single DQ white dwarfs showing a mass distribution peaking below the canonical average ($\approx0.6 M_\odot$) and near $0.5M_\odot$ \citep{cou2019} which suggests that low mass DB (helium-rich) white dwarfs preferentially evolve into DQ white dwarfs \citep{bed2022}.

The carbon abundance in the DQ component of NLTT~16249 indicates the presence of a normal helium envelope $\log{q_{\rm He}}=-3.0$ \citep[see the review by][]{duf2011}. The role of nitrogen as a catalyst in carbon/oxygen production during helium core burning implies its complete elimination by the end of the process. The nitrogen leftovers observed in NLTT~16249 suggest that burning was interrupted before completion, possibly leaving a carbon-dominated core rather than a carbon/oxygen core. The presence of oxygen cannot be confirmed in cool DQ white dwarfs \citep{duf2011}, with limits of C:O$\gtrsim $10:1, although it is found in some of the so-called hot DQ white dwarfs ($T_{\rm eff} \gtrsim12\,000$~K). Ultraviolet spectroscopy offer the only prospect of oxygen detection in NLTT~16249 in the form of CO molecules. Most carbon could actually be unaccounted for based on visual observations alone.

The kinematics of NLTT~16249 informs us on its possible origin.
The radial velocity, $v_r=50.2\pm1.9$~km\,s$^{-1}$, is taken as the weighted average of the apparent velocity of the DA and DQ white dwarfs corrected for their respective  gravitational redshifts. Following \citet{joh1987} the Galactic velocity vector in the local standard of rest with the solar motion from \citet{sch2010} and using the Gaia proper-motion and distance measurements is:
$U=-29.7\pm1.9$, $V=-42.5\pm0.4$, $W=-10.6\pm0.1~{\rm km\,s}^{-1}$.
This inflated ($U,V,W$) vector is typical of an older population. The Galactic orbit integrated with {\tt GALPY} \citep{bov2015} has an eccentricity $e=0.229\pm0.004$ and 
a z-component of the angular momentum $J_z=1436\pm4$~kpc\,km\,s$^{-1}$ characteristic of the thick disc or old thin disc \citep{pau2003,pau2006}. The implied old age, $\tau_{\rm MS}+\tau_{\rm cool}\approx9$~Gyr, would suggest that the progenitors of the double degenerate components lived on the lower (G-type) main-sequence. \citet{kaw2023} investigated the possibility first mentionned by \citet{dun2015} that hot massive DQ white dwarfs may be the results of stellar mergers \cite[see also][]{she2023,kil2024,jew2024}; the double degenerate NLTT16249 may be an example of an object that avoided that fate. In fact, restricting angular momentum losses to gravitational wave emission \citep[see][]{war1995} the system is not expected to merge within $\tau_{\rm merge} \approx 3\times10^{11}$~yr.

\section*{Acknowledgements}

This publication is based on observations collected at the European Organisation for Astronomical Research in the Southern Hemisphere, Chile under program IDs 086.D-0562 and 092.D-0269.
This work has made use of data from the European Space Agency (ESA) mission
{\it Gaia} \footnote{https://www.cosmos.esa.int/gaia}, processed by the {\it Gaia}
Data Processing and Analysis Consortium (DPAC
\footnote{https://www.cosmos.esa.int/web/gaia/dpac/consortium}). 
This publication makes use of data products from the Wide-field Infrared Survey 
Explorer, which is a joint project of the University of California, Los 
Angeles, and the Jet Propulsion Laboratory/California Institute of Technology, 
funded by the National Aeronautics and Space Administration (NASA),
and of data products from the Two Micron All Sky Survey,
which is a joint project of the University of Massachusetts and the Infrared 
Processing and Analysis Center/California Institute of Technology, funded by 
NASA and the National Science 
Foundation. This paper includes data collected by the TESS mission. Funding for the TESS mission is provided by the NASA's Science Mission Directorate. We thank the anonymous referee for many useful suggestions.

\section*{Data Availability}

The VLT/X-shooter and UVES spectra are publicly available at the ESO data archive. \footnote{http://archive.eso.org/cms.html}

\appendix
\section{Molecular band spectra}\label{appenA}

\begin{figure*}
\center
\includegraphics[viewport=20 30 550 550, clip, width=0.6\textwidth]{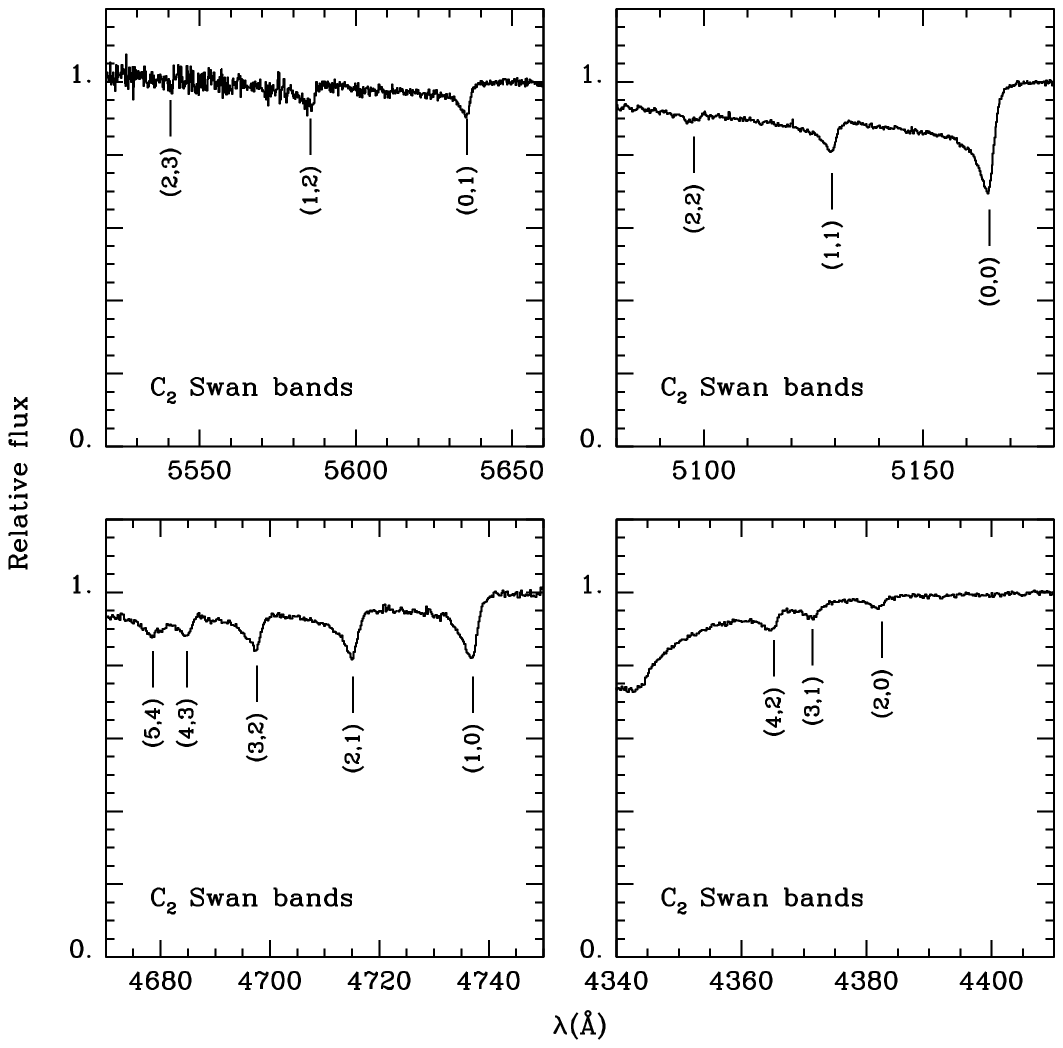}
\caption{C$_2$ Swan bands observed in NLTT~16249 and marked with their respective vibrational numbers ($\nu^\prime,\, \nu^{\prime\prime}$).}
\label{fig-c2swan}
\end{figure*}

\begin{figure*}
\center
\includegraphics[viewport=20 30 550 550, clip, width=0.6\textwidth]{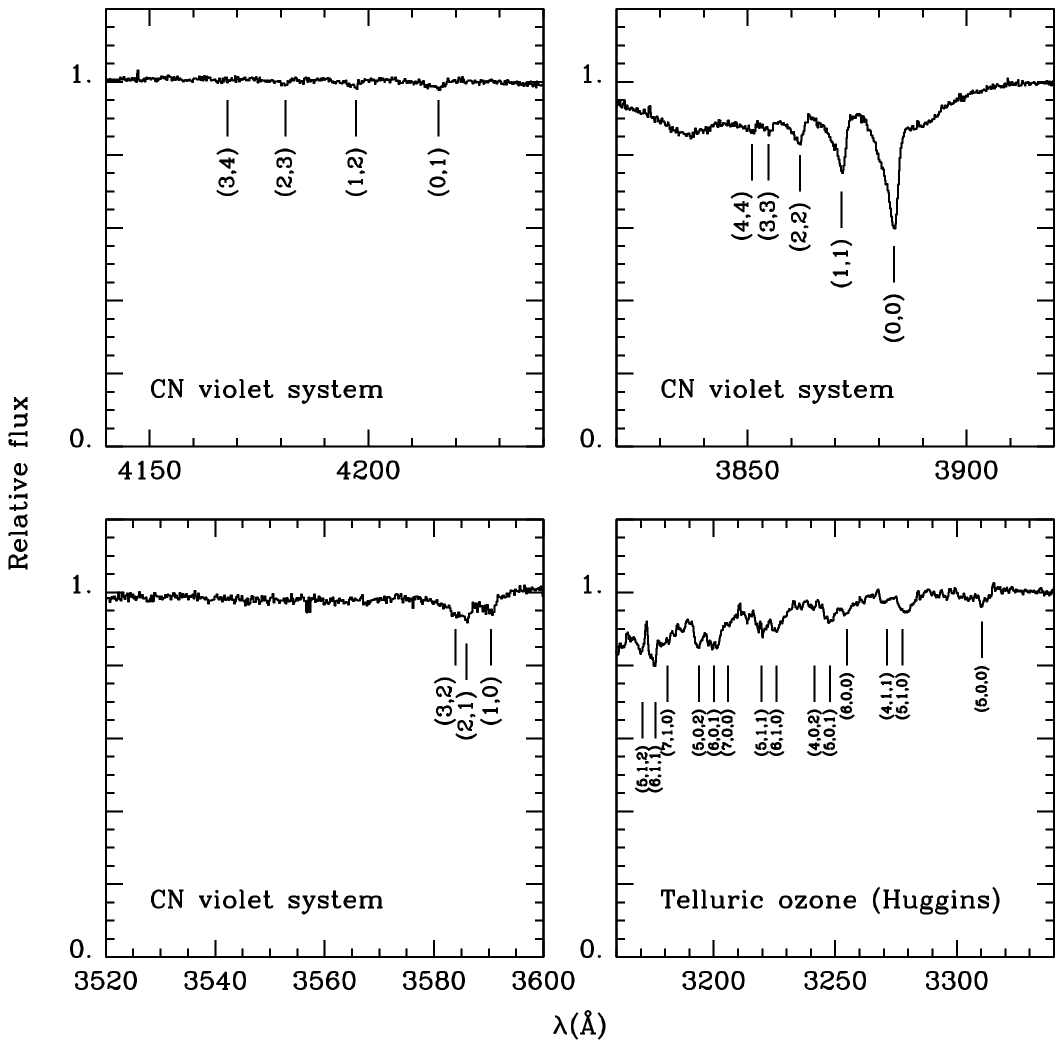}
\caption{CN ultraviolet system bands observed in NLTT~16249 and marked with their respective vibrational transitions ($\nu^\prime,\, \nu^{\prime\prime}$). In addition, Telluric ozone bands (Huggins band) marked with the vibrational numbers ($\nu_1,\nu_2,\nu_3$) are shown in the lower right panel \citep[see][]{leq1992}.}
\label{fig-cnviolet}
\end{figure*}

Figures~\ref{fig-c2swan} and \ref{fig-cnviolet} show details of the C$_2$ Swan and CN violet systems in the X-shooter spectra of NLTT~16249. A line from the C$_2$ Fox-Herzberg system (3283~\AA) could also be present but blended with an ozone line (3277~\AA).

\bsp
\label{lastpage}
\end{document}